\documentclass[fleqn,twoside]{article}
\usepackage{espcrc2}
\usepackage{graphics}
\usepackage{graphicx}
\usepackage[figuresright]{rotating}

\newcommand{\AmS}{{\protect\the\textfont2
  A\kern-.1667em\lower.5ex\hbox{M}\kern-.125emS}}

\title{Confinement and chiral symmetry breaking via a domainlike mean field.}

\author{A.C. Kalloniatis\address{Centre for the Subatomic Structure
          of Matter, University of Adelaide\\ 
        South Australia, 5005, Australia }%
        \thanks{ Supported by the Australian Research Council} 
        S.N. Nedelko\address{Bogoliubov LTPh, JINR\\
      141980  Dubna, Russian Federation}%
        \thanks{Supported by the RFBR grant 01-02-17200}}
\begin{document}

\begin{abstract}
Singular gauge fields in the
partition function for QCD can lead to a domain-like
picture for the QCD vacuum by virtue of constraints 
 on quantum fluctuations at the singularities.
With a simple model of
hyperspherical domains with interiors of constant field strength
we show that the main features of gluon condensation 
and an area law for static quarks can be realised. 
The Dirac operator in such a background is exactly
soluble. Chirality properties of the solutions
show  agreement with recent lattice results.
\vspace{1pc}
\end{abstract}

\maketitle

\section{INTRODUCTION}
The nature and consequences for confinement and chiral-symmetry
breaking of long-range gluon fields in the vacuum have been the
subject of extended investigation.
While many studies rely on one or more of the specific configurations
of monopoles, vortices and instantons, there are arguments
\cite{vBa00} that in fact the complete hierarchy of singular gauge
fields
must play a role, especially if these objects condense in
the vacuum. It is natural then to seek a model in which
the bulk properties of this hierarchy can be effectively
represented. We partially do this here by introducing a model for such
fields in which singularities in vector
potentials are concentrated on three-dimensional hypersurfaces $\partial V_j$
in Euclidean space, in the vicinity of which
gauge fields can be divided into a sum of a singular pure gauge
$S_{\mu}$ and a regular fluctuation part
$Q_{\mu}$, and where a certain $SU(3)_{\rm{colour}}$
colour vector $n_j^a$ can be associated with
$S_{\mu}$.  
Thus 
we have idealised the situation by ignoring lower dimensional 
singular objects which would
appear as "dislocations" at the abelian domain wall
singularities. 

The requirement of finiteness of the action (density) for 
singular background plus fluctuation parts for configurations
dominating the functional integral~\cite{lenz}
can be formalised in our case as specific boundary conditions on $\partial
V_j$ for the fluctuation fields charged with respect to $n_j$.
The interiors of these regions thus constitute ``domains'' $V_j$.
Gauge field modes neutral with respect to $n^a_j$ are not restricted
and generate interactions between domains.

Within the present framework then, we decompose
a general gauge field $A^j_\mu=S^{j}_\mu+Q_\mu^j$
and demanding finiteness  
of the classical action density at the singularity we come to the conditions  
\begin{eqnarray}
\breve n_j Q^{(j)}_\mu  =  0,  \ x\in\partial V_j,
\nonumber
\end{eqnarray}
for gluons, and  
\begin{equation}
\label{bc}
\psi=-i\!\not\!\eta^j e^{i\alpha_j\gamma_5}\psi,
\bar\psi=\bar\psi i\!\not\!\eta^j e^{-i\alpha_j\gamma_5},
x\in\partial V_j
\end{equation} 
for quarks. The adjoint matrix $\breve n_j=T^an_j^a$ appears in the condition 
for gluons, and a bag-like boundary condition arises for quarks,   
with a unit vector $\eta^j_\mu(x)$ normal to $\partial V_j$. 

In a given domain $V_j$ the effect of fluctuations in neighbouring regions
can be seen as an external gauge field $B_{j\mu}^a$ neutral with respect
to $n_j^a$. This enables an approximation  in which
different domains are assumed to be decoupled from each other
but, to compensate, a certain mean field is introduced in domain interiors.
To make the model analytically tractable we consider
spherical domains with fixed radius $R$ and approximate the 
mean field in $V_j$ by a configuration with the field strength
$ \hat {\cal B}_{\mu\nu}^{(j)a} =\hat n^{(j)}B^{(j)}_{\mu\nu}$,
which is (anti-)self-dual $\tilde B^{(j)}_{\mu\nu}=\pm B^{(j)}_{\mu\nu}$
so that $B^{(j)}_{\mu\nu}B^{(j)}_{\rho\nu}=B^2\delta_{\mu\rho}$,
with $B$ a constant the same for all domains. The constant colour matrix  
$\hat n^{(j)}=t^3\cos\xi_j+t^8\sin\xi_j$ belongs to the Cartan
subalgebra with angles $\xi_j\in\{\frac{\pi}{6}(2k+1),\ k=0,\dots,5\}$ 
corresponding to the Weyl subgroup. 
There is no source for this field on the boundary
and therefore it should be treated as strictly homogeneous in all further
calculations. The homogeneity itself appears here just as an approximation.
Thus the partition function we will deal with can be written 
\begin{eqnarray}
{\cal Z} & = & {\cal N}\lim_{V,N\to\infty}
\prod\limits_{i=1}^N\int_V\frac{d^4z_i}{V}
\int\limits_{\Sigma}d\sigma_i
\int {\cal D}Q^i 
\nonumber \\
{}&{}& \times 
{\cal D}\psi_i {\cal D}\bar \psi_i
\delta[D(\breve{\cal B}^{(i)})Q^{(i)}]
\Delta_{\rm FP}[\breve{\cal B}^{(i)},Q^{(i)}]
\nonumber \\
{}&{}& \times 
\exp\left\{
- S_{V_i}^{\rm QCD}
\left[Q^{(i)}+{\cal B}^{(i)},\psi^{(i)},\bar\psi^{(i)}\right]
\right\}
\nonumber
\end{eqnarray}
where the gluons and quarks are integrated over the spaces
${\cal F}^i_Q$ and ${\cal F}^i_\psi$ respectively, and   
the thermodynamic limit assumes $V,N\to\infty$ but 
with the density $v^{-1}=N/V$ taken fixed and finite. 
The measure of integration over parameters characterising domains is defined as
\begin{eqnarray}
\label{measure}
\int\limits_{\Sigma}d\sigma_i\dots & = & \frac{1}{48\pi^2}
\int\limits_0^{2\pi}d\alpha_i
\int\limits_0^{2\pi}d\varphi_i\int_0^\pi d\theta_i\sin\theta_i
\nonumber\\ {}&{}&
\times\int_0^{2\pi} d\xi_i\sum\limits_{l=0,1,2}^{3,4,5}
\delta(\xi_i-\frac{(2l+1)\pi}{6})
\nonumber \\ {}& {}& 
\times\int_0^\pi d\omega_i\sum\limits_{k=0,1}\delta(\omega_i-\pi k)
\dots .
\end{eqnarray}
Here $\varphi_i$ and $\theta_i$ are spherical angles of
the chromomagnetic field,
$\omega_i$ is the angle between the chromomagnetic and
chromoelectric fields, $\xi_i$ is the angle in the colour
matrix $\hat n_i$, 
$\alpha_i$ is the chiral angle
and $z_i$ is the centre of the domain $V_i$.

This partition function describes a statistical system 
of density $N/V$ composed of noninteracting extended
domain-like structures, each of which is
characterised by a set of internal parameters and
whose internal dynamics are represented by the fluctuation fields. 

\section{MEAN FIELD CORRELATORS}  
In this model the connected $n-$point correlator 
of field strength tensors,
\begin{eqnarray}
 B^{a}_{\mu\nu}(x)
=\sum_j^N n^{(j)a}B^{(j)}_{\mu\nu}\theta(1-(x-z_j)^2/R^2),
\nonumber
\end{eqnarray}
can be calculated explicitly using the measure, Eq.~(\ref{measure}). 
Translation-invariant functions
\begin{eqnarray}
\Xi_n(x_1,\dots,x_n)=\frac{1}{v}\int d^4z
\prod_{i=1}^n \theta(1-\frac{(x_i-z)^2}{R^2})
\nonumber
\end{eqnarray}  
emerge and can be seen as the volume of the region of overlap of $n$ 
hyperspheres of radius $R$ and centres ($x_1,\dots,x_n$),
normalised to the volume of a single hypersphere
$v=\pi^2R^4/2$.
The functions $\Xi_n$ are continuous  and 
vanish if $|x_i-x_j|\ge 2R$. 
Correlations in the background field have finite range
$2R$. The Fourier transform of $\Xi_n$ is then an entire analytical 
function and thus correlations do not have a particle interpretation.
The statistical ensemble 
of background fields is not Gaussian since all connected correlators
are independent of each other and cannot be reduced 
to the two-point correlations. 

The simplest application of the above correlators gives  
a gluon condensate density which to this approximation is 
$g^2 \langle F^a_{\mu\nu}(x)F^a_{\mu\nu}(x)\rangle=4B^2.$

Another vacuum parameter which plays
a significant role in the resolution of the $U_A(1)$ problem
is the topological susceptibility 
\cite{Cre77}.
To define this we consider first the topological charge density
$Q(x)={{g^2}\over {32 \pi^2}} \tilde F^a_{\mu\nu}(x)F^a_{\mu\nu}(x)$ 
which to lowest order is
\begin{eqnarray}
Q(x)  ={{B^2}\over {8 \pi^2}} \sum_{j=1}^N\theta[1-(x-z_j)^2/R^2]\cos\omega_j,
\nonumber
\end {eqnarray}
where $\omega_j\in\{0,\pi\}$ depends on the duality of the $j$-th domain.
For a given field configuration the topological charge is additive 
$ Q=\int_V d^4x Q(x)= q(N_+ - N_-)$ for $-Nq\le Q\le Nq$ 
where $q=\frac{B^2R^4}{16}$ is a `unit' topological charge,
namely the absolute value of the topological charge of a single
domain, and $N_{+}$ $(N_-)$ is the number of 
domains with (anti-)self-dual field, $N=N_+ + N_-$. 
The probability  of finding the topological charge $Q$ in a given 
configuration is defined by the 
ratio of the number of configurations  ${\cal N}_N(Q)$ 
with a charge $Q$ 
and the total number of configurations ${\cal N}_N$,
\begin{eqnarray}
{\cal P}_N(Q)=
\frac{N!}{2^N\left(N/2-Q/2q\right)!\left(N/2+Q/2q\right)!}.
\nonumber
\end{eqnarray}
The distribution is symmetric about $Q=0$, where it has a maximum
for $N$ even. 
For $N$ odd the maximum is at $Q=\pm q$. 
Averaged topological charge is zero. 

The topological susceptibility
$\chi$
is determined by the two-point correlator of topological charge density,
which in the lowest approximation gives 
\begin{eqnarray}
\chi = \int d^4x \langle Q(x) Q(0) \rangle 
= {{B^4 R^4} \over {128 \pi^2}}.
\end{eqnarray}

\section{AREA LAW}

To zeroth order in fluctuations the Wilson loop can be written 
\begin{eqnarray}
W(L)&=&\lim_{V,N\to\infty}\prod\limits_{j=1}^N\int_V\frac{d^4z_j}{V}
\int d\sigma_j \nonumber \\
{}&{}&\times \frac{1}{N_c}{\rm Tr}
\exp\left\{i\int_{S_L}d\sigma_{\mu\nu}(x)\hat B_{\mu\nu}(x)\right\}
\nonumber
\end{eqnarray}
where Stokes' theorem has been used. 
Note that path ordering in our case is not necessary since the matrices 
$\hat n^k$ are assumed to be in the Cartan subalgebra.
Computationally it is convenient to
consider a circular contour in the $(x_3,x_4)$
plane of radius $L$ with centre at the origin. 
Details of the calculation are given in \cite{NKYM01}. Put briefly, 
the colour trace is evaluated exactly followed by
the integration over orientations of the vacuum field. 
Integrating over the domain centres 
and taking the thermodynamic limit ($N\to\infty$, $v=V/N=\pi^2R^4/2$),
one obtains for a large Wilson loop $L\gg R$  
\begin{eqnarray}
W(L)=\lim_{N\to\infty}\left[1-\frac{1}{N}U(L)
\right]^N
=e^{-U(L)}   
\nonumber
\end{eqnarray}
with the exponent $U(L)=\sigma \pi L^2 + O(L)$
displaying an area law
with string tension for $SU(3)_{{\rm colour}}$ 
given by $\sigma=Bf(BR^2)$ with the function  
\begin{eqnarray}
\label{sig-su3}
f(z)=\frac{2}{3\pi z}
\left(3-
\frac{\sqrt{3}}{2\pi z}\int_0^{2\pi z/\sqrt{3}}\frac{dx}{x}\sin x
\right. \nonumber \\
\left. - \frac{2\sqrt{3}}{\pi z}\int_0^{\pi z/\sqrt{3}}\frac{dx}{x}\sin x
\right)
\nonumber
\end{eqnarray}
having a purely geometrical origin. 
This function is positive for $z>0$ and has a maximum for $z= 1.55$.
We choose this maximum to estimate the model parameters 
by fitting the string constant to the lattice result,
\begin{eqnarray}
\label{par-val}
\sqrt{B}=947{\rm MeV}, \ R^{-1}=760 {\rm MeV},
\end{eqnarray}
and get for the gluonic parameters of the vacuum
$\sqrt{\sigma}=420 {\rm MeV}, \ \chi=(197 {\rm MeV})^4,
\ \frac{\alpha_s}{\pi}\langle F^2\rangle=0.081({\rm GeV})^4$
and $q=0.15$, while the density of the system is $42 {\rm fm}^{-4}$.
There is no separation of characteristic scales $\sqrt{B}R\approx 1$,
hence an approximation based on large or small domains is
not justifiable and has been not used here. 

If $B$ goes to zero then the string constant vanishes.
This underscores the role of the gluon condensate in the confinement of 
static charges.  On the other hand, if the number of domains is fixed and 
the thermodynamic limit is 
defined as $V,R\to\infty, N={\rm const.}<\infty$, namely if  
the domains are macroscopically large,
then $W(L)=1$, which indicates the absence of a linear potential
between infinitely massive charges in a purely homogeneous field. 

 An area law obviously does not occur for adjoint charges 
because of  the presence of 
zero eigenvalues of the adjoint matrix $n^aT^a$.

\section{DYNAMICAL CONFINEMENT}

Confinement of the fluctuation fields can be seen in the  
analytical properties of their propagators. 
For the above boundary conditions
the fluctuation quark and gluon propagators 
can be analytically calculated by reduction 
to the charged scalar field problem. 
This in turn is essentially just that  
of a four-dimensional harmonic oscillator with the 
orbital momentum coupled to the external field, and
the general solution has been found exactly by 
decomposition over hyperspherical harmonics
\cite{NKYM01}.

Due to the above boundary conditions 
the $x-$space propagators of charged fields are defined in
regions of finite support where they have integrable
singularities so that their Fourier transforms 
are entire functions in the complex momentum plane.
This is consistent with a confinement of dynamical fields\cite{leutw}. 
Entire propagators have the physically appealing property
of a Regge spectrum of relativistic bound states\cite{efi}.
 
\section{CHIRALITY OF QUARK MODES}
Turning to quarks specifically, we address the eigenvalue problem 
for the massless Dirac operator in a domain,
\begin{eqnarray}
&&(i\!\not\!D-\lambda)\psi(x)=0,
\nonumber
\\
&&D_\mu=\partial_\mu-i\hat B_\mu
=\partial_\mu + \frac{i}{2}\hat n B_{\mu\nu}x_\nu,
\nonumber
\end{eqnarray}
subject to the boundary condition Eq.(\ref{bc}). 
We introduce projectors in the colour direction
$ N_\pm=\frac{1}{2}(1\pm \hat n/|\hat n|)$
and spin projection with respect to the magnetic field
$\Sigma_\pm=\frac{1}{2}(1\pm \vec\Sigma\vec B/B)$
with $\hat B=|\hat n|B$.
It is also useful to introduce the mixed projector
$O_{\zeta}=N_+\Sigma_{\zeta} + N_-\Sigma_{-\zeta}$.
Substituting 
\begin{eqnarray}
\psi & = &(i\!\not\!D+\lambda)\Phi(x),
\nonumber \\
\Phi&=&P_\pm \Phi_0+P_\mp O_+\Phi_{+1}+ P_\mp O_-\Phi_{-1}
\label{q-pr1}
\end{eqnarray}
shows that $\Phi_\zeta$ satisfies  
\begin{eqnarray}
(-D^2+2\zeta\hat B +\lambda^2)\Phi_\zeta=0,
\nonumber
\end{eqnarray}
whose exact solution is given in \cite{NKYM01}.
It turns out that it is impossible to fulfill the boundary
condition Eq.(\ref{bc}) with the term $\Phi_0$ in Eq.(\ref{q-pr1}).
Thus $\Phi$ must have definite chirality corresponding to the
duality of the vacuum gluon field.  

The boundary condition enforces a discretisation of the
eigenvalues $\lambda$. 
The states are then labelled by
the set of quantum numbers: ${\cal N} = n,k,m_1,m_2,\zeta, s$ with
$n=0,1,\dots $ a radial quantum number labelling the solutions of the
boundary condition for given $O(4)=O(3)\times O(3)$ angular quantum numbers 
$k,m_1,m_2$, $\zeta = \pm 1$ corresponding
to the colour projection with respect to $\hat{n}$ and $s=\uparrow, \downarrow$
representing the spin projection with respect to the magnetic field.
Zero modes are absent under chosen boundary conditions.
The  eigenspinors have the structure
\begin{eqnarray}
\psi_{\cal N}=i\!\!\not\!\eta \chi_{\cal N} +\varphi_{\cal N}, 
\gamma_5\chi_{\cal N}=\mp\chi_{\cal N}, 
\gamma_5\varphi_{\cal N}=\mp\varphi_{\cal N},
\nonumber
\end{eqnarray} 
 where spinors $\varphi_{\cal N}$ and $\chi_{\cal N}$ 
as well as  the constraints defining
the corresponding eigenvalues are known in analytical form and will be
discussed in detail elsewhere~\cite{{NKqu01}}. 
For illustration we give
the eigenvalue equations for the purely radial $(k=m_1=m_2=0)$ modes. With
$\Lambda=\lambda/\sqrt{2\hat B}$, $z=\hat Br^2/2$ we have
the equation for $\Lambda_{n000}^{\downarrow+}$ 
\begin{eqnarray}
\sqrt{z}\Lambda
M\left(\Lambda^2+1,3,z\right)
+2e^{\mp i\alpha}M\left(\Lambda^2,2,z\right)=0
\nonumber 
\end{eqnarray}
and that for $\Lambda_{n000}^{\downarrow-}$ 
\begin{eqnarray}
\sqrt{z}
\left[\frac{\Lambda^2+2}{2}
M\left(\Lambda^2+3,3,z\right)
- M\left(\Lambda^2+2,2,z\right)\right]
\nonumber \\
+e^{\mp i\alpha}\Lambda M\left(\Lambda^2+2,2,z\right)=0
\nonumber 
\end{eqnarray}
and $\Lambda_{n000}^{\uparrow+} = \Lambda_{n000}^{\downarrow-}$ and
$ \Lambda_{n000}^{\uparrow-} = \Lambda_{n000}^{\downarrow+}$.
Here $M(a,b,z)$ is the confluent hypergeometric function.
The sign minus (plus) in front of $i\alpha$ corresponds to the 
(anti-)self-dual domain.

With these solutions we investigate the chirality of the eigenmodes and 
also how their probability density correlates with the underlying domain.
Following \cite{Hor} we introduce the local chirality parameter
$X(x)$ defined  via
\begin{eqnarray}
\tan \left( \frac{\pi}{4}(1 + X(x)) \right)
= {{|\psi_L(x)|} \over {|\psi_R(x)|} },
\nonumber
\end{eqnarray}
which will give extremal values $X=\pm 1$  at positions
$x$ where $\psi(x)$ is purely right(left) handed. 
For lattice overlap fermions, even in uncooled
gauge fields, histograms of $X$ measured at  
probability density maxima for low-lying overlap-Dirac eigenmodes 
show peaks close to the extremal values \cite{EH01}. 
This indicates that low-lying  modes are strongly chiral and are thus   
a useful filter for the duality of ``objects'' underlying the
gauge field fluctuations. 
%%%%%%%%
\begin{figure}[htb]
\includegraphics*{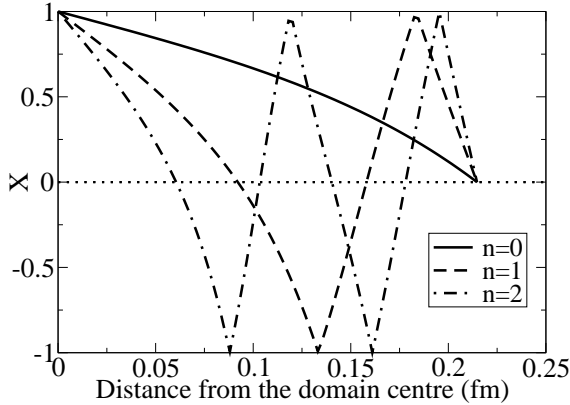}
\vspace*{-14mm}
\caption{Chirality parameter for the lowest radial modes
$\psi_{n000}^{\downarrow+(\uparrow-)}$, self-dual domain, $\alpha=\pi/2$. }
\label{fig:X}
\vspace*{-6mm}
\end{figure}
%%%%
Within the present model we observe that at the centres of
domains all of the eigenmodes are exactly chiral and probability densities
are maximal. The ``width'' of the peaks for the lowest 
modes at half-maximum varies for different values $\alpha$ and  is of the order of
$.12-0.14 {\rm{fm}}$, consistent with the lattice observations of \cite{Hor2}.
As is  seen from Fig.(\ref{fig:X}) the chirality of the lowest mode ($n=0$)
monotonically decreases with distance from the centre.
The chirality parameter for the excited modes alternates  
between extremal values,
the number of alternations is correlated with the radial number $n$, and the 
half-width decreases with growing $n$. The chirality parameter
is zero at the boundary for all modes. 
Qualitatively this picture does not depend on angle $\alpha$.

\begin{figure}[htb]
\includegraphics*{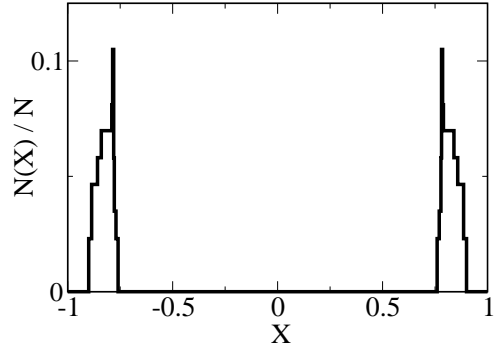}
\vspace*{-10mm}
\caption{Histogram of chirality parameter $X$
averaged over the central region with radius $0.05$fm.}
\label{fig:X-hist}
\vspace*{-8mm}
\end{figure}

The degree of chirality of the modes can be 
conveniently characterized via a histogram  
if we average X(x) over a small neighborhood of the domain centre.
Then, by assuming that all values of $\alpha$
are equally probable  we compute the probability to find a given value
of smeared $X$ among a set of modes. The result given in Fig.(\ref{fig:X-hist})
was obtained for the lowest modes with all possible spin-color orientaions. 
We observe the
double peaking close to the extrema with $X\approx \pm 0.8$. 
Including higher modes will broaden the peaks and build up a central plateau.    
This feature as well as the above mentioned values for the half-decay and 
the density of domains is 
in qualitative and quantitative agreement with 
recent lattice results \cite{EH01,Hor2}.  However, at this stage, it would be
premature to go beyond purely numerical comparison of our results
and those of the lattice. 
%%%%%%%%
%%%%%
Finally, we quote the estimate of the quark condensate density
averaged over (anti-)self-dual configurations and $\alpha$
obtained in~\cite{NKYM01} directly from the quark propagator.  
For  $B$ and $R$  as in Eq.~(\ref{par-val})
the density at the domain  center
is equal to
$ \langle \bar\psi \psi\rangle_{\rm center}= -(228 {\rm MeV})^3. $
It should be stressed that this nonzero result is 
due to the non-integral topological charge $q$ associated with domains.
This and our results for the chirality parameter $X$
suggest that the underlying mechanism of chiral symmetry breaking
here is the strong chirality of the low-lying non-zero modes,
which in turn is a consequence of the (anti-)self-duality of
the background gluonic configurations and their non-integral topological charge.


\begin{thebibliography}{9}

\bibitem{vBa00} P. van Baal, hep-ph/0008206, 
At the frontier of particle physics, vol. 2 683-760,
ed. by M. Shifman, World Scientific, 2001. 
\bibitem{lenz} F. Lenz, O. Jahn, Phys. Rev. D58 (1998) 085006; 
F. Lenz, S. Worlen, hep-ph/0010099,
At the frontier of particle physics, vol. 2 761-821,
ed. by M. Shifman, World Scientific, 2001.
\bibitem{Cre77}
R.J. Crewther, Phys.Lett. B70 (1977) 349; 
G. Veneziano, Nucl. Phys. B159 (1979) 213; 
E. Witten,  Nucl. Phys. B156 (1979) 269.
\bibitem{NKYM01} A.C.Kalloniatis, S.N. Nedelko,
Phys. Rev. D64 (2001) 114025.
\bibitem{leutw}H. Leutwyler, Phys. Lett. B96 (1980) 154.
\bibitem{efi} G.V. Efimov, S.N. Nedelko, 
Phys. Rev. D51 (1995) 174; 
G.V. Efimov, G. Ganbold, hep-ph/0103101.
\bibitem{NKqu01} A.C. Kalloniatis, S.N. Nedelko, {\it in preparation}.
\bibitem{Hor} I.~Horvath et al., hep-lat/0102003. 
\bibitem{EH01} R.G.~Edwards, U.M.~Heller, hep-lat/0105004.
\bibitem{Hor2} S.J. Dong et al., hep-lat/0110037.  
\bibitem{wipf} S. Duerr, A. Wipf, Nucl. Phys. B443 (1995) 201.
\end{thebibliography}
\end{document}